# Cefazolin versus anti-staphylococcal penicillins for treatment of methicillin-susceptible *Staphylococcus aureus* bacteremia: a narrative review.

Paul Loubet[1,2], Charles Burdet[1,3], William Vindrios[2], Nathalie Grall[1,4], Michel Wolff[1,5], Yazdan Yazdanpanah[1,2], Antoine Andremont[1,4], Xavier Duval[1,6], François-Xavier Lescure[1,2]

1. IAME, UMR 1137, INSERM, Université Paris Diderot, Sorbonne Paris Cité, Paris, France
2. AP-HP, Hôpital Bichat-Claude Bernard, Service de Maladies Infectieuses et Tropicales, Paris, France
3. AP-HP, Hôpital Bichat-Claude Bernard, Département d'épidémiologie, biostatistique et recherche clinique, Paris, France
4. AP-HP, Hôpital Bichat-Claude Bernard, Laboratoire de Bactériologie, Paris, France
5. AP-HP, Hôpital Bichat-Claude Bernard, Service de réanimation médicale et infectieuse, Paris, France
6. AP-HP, Hôpital Bichat-Claude Bernard, Centre d'Investigation Clinique, Paris, France

**Corresponding author**:
Dr François-Xavier Lescure
Service de Maladies Infectieuses et Tropicales
Hôpital Bichat-Claude Bernard
46 Rue Henri Huchard. 75018 Paris, France
Tel: + 33 1 40 25 78 03
Fax: + 33 1 40 25 88 60
Electronic address: xavier.lescure@aphp.fr

**Abstract.**

**Background**. Anti-staphylococcal penicillins (ASPs) are recommended as first-line agents in methicillin-susceptible *Staphylococcus aureus* (MSSA) bacteremia. Concerns about the safety profile have contributed to the increased use of cefazolin. The comparative clinical effectiveness and safety profile of cefazolin versus ASPs for such infections remain unclear. Furthermore, uncertainty persists concerning the use of cefazolin due to controversies over its efficacy in deep MSSA infections and its possible negative ecological impact.

**Aims.** The aim of this narrative review was to gather and balance available data on the efficacy and safety of cefazolin versus ASPs in the treatment of MSSA bacteremia and to discuss the potential negative ecological impact of cefazolin.

**Sources**. PubMed and EMBASE electronic databases were searched up to May 2017 in order to retrieve available studies on the topic.

**Contents**. While described in vitro and in experimental studies, the clinical relevance of the inoculum effect during cefazolin treatment of deep MSSA infections remains unclear. It appears that there is no significant difference in rate of relapse or mortality between ASPs and cefazolin for the treatment of MSSA bacteremia but these results should be cautiously interpreted because of the several limitations of the available studies. Compared to cefazolin, there is more frequent discontinuation for adverse effects with ASPs use, especially because of cutaneous and renal events. No study has evidenced any change in the gut microbiota after the use of cefazolin.

**Implications**. Based on currently available studies, there is no data allowing to choose one antibiotic over the other except in patients with allergy or renal impairment. This review points out the need for future prospective studies and randomized controlled trials to better address these questions.

1. **Background**

Methicillin-susceptible *Staphylococcus aureus* (*S. aureus*) (MSSA) bacteremia remains a major cause of community- or hospital-acquired bloodstream infections, with approximately 200,000 cases occurring annually in Europe [1] and high in-hospital mortality (25-35%) [2–4]. Anti-staphylococcal penicillins (ASPs) such as oxacillin or cloxacillin are recommended as first-line agents, but their use may be limited by concerns about their safety profile and difficult dosing schedule in patients with renal failure. Cefazolin, an intravenous first-generation cephalosporin, (1GC) is thus more and more used as an alternative option [5,6]. Yet the comparative clinical effectiveness and safety profile of cefazolin versus ASPs (nafcillin, oxacillin, cloxacillin, dicloxacillin or flucloxacillin) for MSSA infections remain unclear because of the limited published data. Furthermore, uncertainty persists for the curative use of cefazolin because of controversies concerning its efficacy in high-inoculum deep MSSA infections and its possible negative ecological impact.

2. **Aims**

The aim of this narrative review was to gather and balance available data on the efficacy and safety of cefazolin versus ASPs in the treatment of MSSA bacteremia and to discuss the potential negative ecological impact of cefazolin.

3. **Sources**

An extensive search of PubMed (January, 1985, to May, 2017) and EMBASE (January, 2010, to May, 2017) was performed to identify relevant studies for our review. Search terms included "cefazolin", "oxacillin", "nafcillin", "antistaphyloccocal penicillin", "methicillin", "β-lactams", "bacteremia", "bacteraemia", "bloodstream infection", "efficacy", "safety", "effectiveness", "inoculum effect", "gut microbiota", "resistances", "*Staphylococcus aureus*", and "MSSA". The reference lists of all articles retrieved were checked for additional relevant references. Two reviewers (PL and FXL) independently searched the literature and examined relevant studies. A study was considered eligible if the role of cefazolin in comparison with an anti-staphylococcal penicillins in the treatment of infections caused by methicillin-susceptible *Staphylococcus aureus* was assessed. Furthermore, clinical or experimental studies dealing with the existence of an inoculum effect with the use of cefazolin and occurrence of changes in gut microbiota following cefazolin and ASP use were included. Only studies published in English were considered in this review.

## 4. Content
### a. Inoculum effect

The inoculum effect has been defined as a significant rise in the cefazolin minimum inhibitory concentration (MIC) when the bacterial inoculum size is increased to $10^7$ colony-forming units (CFU)/mL (instead of the standard $10^5$ CFU/mL) [7]. Four different types (A, B, C and D) of staphylococcal β-lactamase enzymes have been characterized based on their substrate specificity and amino acid sequence [8], and each of these has a different substrate profile [9]. An inoculum effect of β-lactamase in MSSA has been suggested in vitro, with an MIC increase especially with *blaZ* type A β-lactamase. Type A β-lactamase efficiently hydrolyzes cefazolin [10], but, not all isolates producing type A β-lactamase exhibit a significant cefazolin inoculum effect [11–14] because of mechanisms that are not clearly known [15]. However, a recent study has suggested that there might exist an association between type A *blaZ* gene polymorphism and cefazolin inoculum effect [16].

In our review, the prevalence of β-lactamase ranges from 77 to 92% while Type A represents 15-34% and a cefazolin inoculum effect was found in 13 to 58% of the MSSA isolates. It seems that no significant association exists between inoculum effect positivity and demographic factors, underlying disease or site of infection [17] even though one study found that osteomyelitis is highly associated with cefazolin inoculum effect in South American hospitals [14]. Five studies assessed the clinical outcomes of the patients from whom MSSA isolates were collected depending on the presence of an inoculum effect. None found an impact of inoculum effect on mortality at day 90 and/or treatment failure. However, none of the studies was powered enough to evaluate the clinical impact of the cefazolin inoculum effect and the blaZ gene type. (**Table 1**)

In vivo results are conflicting (**Table 2**). Studies of MSSA infective endocarditis have shown that the in vitro inoculum effect may have consequences [18–20], whereas other studies suggest that the slow inactivation of cefazolin by staphylococcal β-lactamase is of little importance, because diffusion into the area of infection occurs rapidly enough to yield effective antibacterial concentrations [21].

In conclusion, the hydrolysis of cefazolin by *S. aureus* type A β-lactamases in high-inoculum deep infections has been proven in vitro. However, its frequency in MSSA bacteremia has been found to be limited, ranging from 13 to 58% [12–14,17,22,23]. The fact that it may lead to potential therapeutic failures is still debated with conflicting results in animal studies and six human studies that found no impact of the inoculum effect. However, these studies are limited by their small sample size, low rate of deep-seated infections and the presence of selection bias. Furthermore, the fact that neither susceptibility testing for cefazolin for MSSA nor the presence of type A β-lactamases are routinely tested makes it difficult to gather data on the topic and to establish practical recommendations.

    **b. Clinical efficacy (Table 3)**

Cloxacillin and cefazolin are more effective in the treatment of MSSA bacteremia than alternative treatments, with 2-fold and 3-fold increases in mortality rate with other β-lactams [24] and vancomycin [25–29], respectively.

Although ASPs are the recommended treatment in MSSA bacteremia, the use of cefazolin in increasing. However, the quantity and quality of publishing data comparing clinical effectiveness of cefazolin versus ASPs are limited. So far, seven observational studies have compared cefazolin to ASPs in the treatment of MSSA bacteremia [24,30–35]. Six of these studies found no difference in treatment failure and/or mortality between cefazolin and ASPs groups with half of the studies reporting that cefazolin was associated with non-significant lower mortality [32–34]. The more recent study from McDanel *et al.*, which is the largest one with 1163 patients in the cefazolin group and 2004 patients in the ASPs group found that cefazolin was significantly associated with lower mortality (aHR 0.8 [0.7-0.9]).

Details on patients and infections characteristics, antibiotics dosing and duration, outcomes of interest and methods to control bias in the seven studies are displayed in **Table 1**.

All of these studies have several limitations. First, they all are retrospective with small sample size (except for McDanel *et al.*). Second, important data such as, type and duration of empiric therapy before the start of cefazolin or ASPs, duration of bacteremia, antibiotics dosing, rates of metastatic infection and source control, are often missing. Third, the rate of deep seated infections, defined as endocarditis, bone or joint infection, device related infection, deep-seated abscess and pneumonia is relatively small. Fourth, while ASP dosing was the same across the studies, cefazolin dosing ranged from 3g/day to 6g/day making comparison difficult between studies. Finally, despite statistical adjustments to allow better comparison between groups, these studies are facing selection biases with imbalances between study groups with more severe, including more deep-seated and metastatic, infections and less source control in the ASP groups. Results of studies showing a better efficacy of cefazolin may be partly explained by the fact that cefazolin was used in less severe patients mostly with catheter or skin and soft tissue related bacteremia with easier source control and thus should be cautiously interpreted.

### c. Safety (Table 4)

Patients with MSSA infection often require prolonged administration of high-dose parenteral antimicrobial therapy with standard doses of 12 grams per day for oxacillin (25 to 50 mg/kg/4 to 6 hours) and 6 grams per day for cefazolin (25 to 50 mg/kg/8h). Because of aging and cumulating comorbidities in these patients, safety issues following the use of ASPs are not infrequent, especially hypersensitivity reactions (more than 10%) [36,37] and renal impairment (more than 10%) [38]. Premature discontinuation of ASPs attributed to adverse events have been reported in 17 to 21% of treated patients for complicated MSSA bacteremia with standard doses of oxacillin or nafcillin (12g/24h) [30,31]. In the case of chronic kidney disease with decreased glomerular filtration rate, ASP dosing is not clearly known.

Five studies have compared the occurrence of adverse events between cefazolin and ASPs among patients treated for MSSA bacteremia [30,31,33,38,39]. All but one study report higher adverse drug events in ASPs groups mainly due to nephrotoxicity and hypersensitivity reactions. These adverse drug events often required antibiotics discontinuation.

It appears that adverse event and criteria for discontinuation are not clearly defined across the studies with a wide range of nephrotoxicity definition for example. Furthermore, due to the retrospective nature of these studies, the quality of data collection is poor with important information biases. As for all observational studies, selection biases affect these safety studies. Severe patients, more likely to be concerned by acute renal failure or overdosing, are more frequently treated with ASPs.

### d. Ecological impact on gut microbiota

In the current context of growing bacterial resistance, especially 3GC-resistant Enterobacteriaceae, the potential negative gut ecological impact of cephalosporins compared to very narrow spectrum antibiotics such as ASPs is largely debated. From a theoretical point of view, high biliary excretion of antibiotics and a sparing spectrum for anaerobes and lactobacilli may foster selection of high MIC Enterobacteriaceae, Clostridia and Candida by influencing the ecological balance of the gut microbiota. The biliary excretion of cefazolin is low and amounted to 0.03% of the administered dose, while 2–10% of a dose of cloxacillin or oxacillin can be recovered from bile [40].

### i. Ecological effect of ASPs on gut microbiota

Although, there are many studies of the effects of ASPs on skin flora, data on the ecological effects of ASPs on gut microbiota are scarce. Narrow spectrum penicillins seem to present a low risk for diarrhea associated with *C. difficile* in a systematic literature review published in 1998; ASPs seemed to present one of the lowest risks (OR 3.2; 95% CI 1.7–6.2), close to that of vancomycin (3.1; 95% CI 1.8–5.2), and much lower than broad spectrum antibiotics such as amoxicillin/clavulanic acid combination for example (22.1; 95% CI 6.5–75.4) [41].

A Japanese study assessed the effect of antibiotics on the fecal flora in hospitalized children aged from 1 to 12 years, who received ampicillin (n=6), methicillin (n=8), cefpiramide (n=7) or ceftazidime (n=7). Antibiotic use was given for 5 to 14 days. Fifteen same aged hospitalized children who did not receive any antimicrobials served as controls. There was no significant decrease in the count of *Enterobacteriaceae* in patients treated with methicillin [42].

### ii. Ecological effect of C1G on gut microbiota

Ambrose *et al*. studied the influence of a single intravenous dose of antibiotic on gut microbiota and the emergence of *C. difficile* over two weeks in 78 volunteers (13 groups of 6 volunteers). Each group of 6 received either penicillins, from among benzyl penicillin, ampicillin, mezlocillin, piperacillin, ticarcillin, or cephalosporins, from among 1CG to 3CG, and the results were compared with those for a control group of 6 volunteers who received no antibiotic. Only cephalosporins were found to be associated with emergence of *C. difficile*, penicillins and controls were not. When they considered total aerobic counts, only the reduction after ceftriaxone achieved statistical significance (P<0.025) with decrease in counts of *Escherichia coli*, though an increase in the counts of *enterococci* was also observed in all groups. For the anaerobe count, only cefotetan was associated with a trend for decrease. Overall, no significant changes were observed for cefazolin [43].

Knothe *et al*. investigated in healthy volunteers the effects on gut microbiota of 1GC (cefazolin) and 3CG (cefotaxime). One or two stool specimens were taken before, during and several days after medication. No selection of strains resistant to ampicillin or cefazolin occurred, while cefazolin considerably reduced *Bacteroides* spp., *lactobacilli* and *Enterobacteriaceae* [44].

Finally, Takesue *et al*. investigated changes in gut microbiota in 24 patients given intravenous antibiotics for a 4-day period after gastrectomy. Patients were divided into 3 groups with 1CG (cefazolin), 2CG (flomoxef) and 4GC (cefozopran). Cefazolin had less of an effect on the gut microbiota changes. Flomoxef caused the most remarkable change in anaerobic bacteria while the number of *Enterobacteriaceae* decreased significantly only with 4CG. [45].

While, the negative ecological impact of cephalosporin use is known, few clinical studies specifically assessed the impact of 1GC on gut microbiota [43–45]. Indeed, despite its wide use in antimicrobial prophylaxis in surgical care, no study has clearly assessed the ecological impact of cefazolin. In the reviewed studies, no change in counts of *Enterobacteriaceae*, *enterococci*, yeasts, total anaerobes, *Clostridia* spp. or *Bacteroides* spp. was observed after administration of cefaloridine, cefalotine or cefazolin. However, it seems that significant changes begin from second cephalosporin generations use [43].

In the current era of growing antimicrobial resistance, the ecological impact has to be considered among potential adverse effects of antibiotics, especially when one has to balance between penicillin and cephalosporin. The more appropriate populations to assess gut microbiota dysbiosis under antimicrobial are healthy volunteers as well as patients undergoing surgery and receiving antimicrobial prophylaxis. Unfortunately, too few studies have been done in these populations. Furthermore, judgement criteria used in available studies are not accurate enough to conclude. Finally, it has to be underlined that none of the studies assessed emergence of 3GC resistant *Enterobacteriaceae* and that genetic sequencing methods have not been used to analyze stools. Waiting for such data, cefazolin appears to have a very limited gut ecological impact.

5. **Implications**

The quality of the published data comparing ASPs and cefazolin as treatment options for MSSA bacteremia is insufficient while the associated morbidity and mortality are high in this frequent disease. While described in vitro and in experimental studies, the clinical relevance of the inoculum effect during cefazolin treatment of deep MSSA infections remains uncertain. This inoculum effect which appears to be infrequent, is not routinely tested in microbiological labs making its impact difficult to assess in routine care.

From a clinical point of view, it seems that there is no difference in efficacy between these drugs. However, available data on clinical efficacy are from retrospective studies that are affected by selection biases issue. Despite concerns about the possible negative ecological impact of cefazolin, no studies have evidenced changes in gut microbiota after its use, but the designs of the available studies are all too old to be able to correctly assess this issue. Concerning safety, it appears that adverse events, especially cutaneous and renal, are more frequent with ASPs than with cefazolin. All these points need to be confirmed in randomized controlled trials that should take into account ecological data.

Based on these reviewed data and our clinical experience, we suggest using cefazolin in catheter related infections, skin/soft tissue infection, non-complicated IE and bone and joint infection because of the excellent bone penetration [46,47]. Conversely, because of the poor penetration of cefazolin through the blood-brain barrier [48,49], ASPs should be preferred for central nervous system infections. In case of complicated IE and deep-seated abscesses, because of the hypothetical risk of clinical failure due to the inoculum effect, ASPs should rather be considered along with source control when possible. In case of deep-seated infection and complete stock-out of ASPs, source control and an increase in cefazolin dosing (>6g/day) should help mitigating the inoculum effect.

**Competing interests**.

The authors declare no commercial or other association that might pose a conflict of interest.

**Funding statement**.

The current work received no funds.

**Table 1. Characteristics and results of studies assessing in vitro and in vivo inoculum effect in humans**

| | Study Design | Number of MSSA bacteremia | Infection sites | Prevalence of BlaZ (%)* | Type | Definition of IE | Prevalence of IE (%) | Comment | Clinical Outcomes¶ | Number of patient with IE | Number of patient without IE | Results |
|---|---|---|---|---|---|---|---|---|---|---|---|---|
| **Nannini et al. 2009** | Multicenter Retrospective study | 98 | Endocarditis 30% Hospital acquired Pneumonia 30% Skin and soft tissue 29% Unknown 11% | 87 | Type A 26% Type B 15% Type C 46% | MIC >16 µg/mL with $10^7$ CFU/mL | 19 | At high inoculum, type A producers displayed higher cefazolin MICs than type B or C producers (11.2 vs. 2.8 (p=0.002) and 5.6 µg/mL (p=0.04) respectively) | Treatment failure | 3 | 9 | 3 (100%) vs 3 (33%) p=0.2 |
| **Livorsi et al. 2012** | Multicenter Retrospective study | 185 | Unknown 46% Bone and joint 16% Catheter related 14% Endocarditis 9% Pneumonia 7% | 77 | Type A 34% Type B 30% Type C 35% Type D 1% | ≥4-fold increase in MIC from a standard to a high inoculum | 27 | 4% of isolates (8/185), all type A Bla strains, demonstrated a non-susceptible cefazolin MIC. | D90 Treatment failure | 2 | 5 | No significant differences between the two groups |
| **Rincon et al 2013** | Multicenter Retrospective study | 296 | NA | NA | Type A 67% Type C 29% | MIC >16 µg/mL with $10^7$ CFU/mL | 33 | - | NA | NA | NA | - |
| **Chong et al. 2014** | Single center Retrospective study | 220 | NA | 92 | Type A 17% Type B 20% Type C 53% | ≥4-fold increase in MIC from a standard to a high inoculum | 13 | - | Relapse of infection D90 Mortality Treatment failure | 10 | 67 | 0 (0%) vs 2(4%) p=1 2 (20%) vs 12 (18%) p=1 0 (0%) vs 10 (15%) p=0.34 |
| **Lee et al. 2014** | Multicenter Retrospective study | 113 | Skin and soft tissue 35% Unknown 20% Catheter related 18% | 78 | Type A 15% Type C 41% | ≥4-fold increase in MIC from a standard to a high inoculum | 58 | - | Treatment failure | 65 | 48 | IE was not associated with treatment failure (aOR 1.3 95%CI 0.4-4.9, p=0.7) |
| **Song et al. 2014** | Multicenter Retrospective study | 303 | Bone and joint 25% Skin and soft tissue 22% Unknown 21% Pneumonia 14% Catheter related 10% | 84 | Type A 13% Type B 27% Type C 44% Type D 0.3% | MIC >16 µg/mL with $10^7$ CFU/mL | 20 | CIE positivity was found to be significantly associated with the type of the blaZ gene. 56% (23/41) in Type A; 28% (37/132) in Type C and 1.2% (1/81) in Type B) | D90 Mortality | 61 | 242 | IE was not associated with treatment failure (OR 1.7 95%CI 0.9-3.3, p=0.13)£ |

MSSA: MSSA= Methicillin-Susceptible *Staphylococcus*, IE: Inoculum Effect
*Percentage of strains producing Beta lactamase (Bla)
¶**In patients with strains producing Bla and receiving cefazolin as a definitive therapy.**
£Univariate analysis

**Table 2. Characteristics and results of studies assessing in vitro and vivo inoculum effect in animals**

| | Model | Antibiotics compared | In vitro | | In vivo | |
|---|---|---|---|---|---|---|
| | | | Outcomes | Results | Outcomes | Results |
| **Carrizosa et al. 1979** | Experimental endocarditis in rabbits. Highly penicillin-resistant strain of *S. aureus*. | Cefazolin Cephalotin Methicillin | Mean log10 CFU (standard deviation) per gram of vegetation after 3 days of therapy | Cefazolin every 6 hours: 4.2 ± 2.7 Cefazolin every 8 hours: 4.7 ± 2.8 Cephalothin: 4.5 ± 2.7 Methicillin: 2.0 ± 0 ($P < 0.01$ for comparisons between methicillin and each of the three other groups). | Positivity of Blood cultures* | Methicillin 0/10 (0%) Cefazolin (every 6 hours) 2/13 (15%) Cefazolin (every 8 hours) 2/13 (15%) Cephalothin: 3/13 (23%) |
| **Goldman and Petersdorf 1980** | Experimental endocarditis in rabbits. Beta-lactamase-positive vs. Beta lactamase negative strains | Cefzaolin Cephalothin (1GC) Cefoxitin (2GC) Cephaloridin (1GC) | Minimum Inhibitory Concentration (MIC) | **In β-lactamase-producing strain** The MIC of cefazolin was increased with *S. aureus* inoculum, unlike other cephalosporins **In β-lactamase-negative strain**, MICs were stable for all antibiotics | Day 4 mortality Cefazolin vs Cephalotin | 65% (13/20) vs 20% (4/20) ($p<0.005$) |
| **Kaye et al. 1979** | Artificial intraperitoneal infection in rabbits Highly penicillin-resistant strain. | Cefazolin Cephalothin (1GC) Cefoxitin (2GC) Cefamandole (1GC) | S. aureus bacterial counts in intraperitoneal infection after 8 days of treatment (CFU/mL) | Cefazolin: 1.6 log10, Cephalotin: 1.0 log10 Cephoxitin: 1.9 log10 Cefamandole: 3.6 log10 | - | NA |
| **Nannini et al. 2013** | Experimental endocarditis in rats. Type A β-lactamase producer strains (TX0117) vs. standard strains (TX0117c) | Cefazolin Nafcillin Daptomycin | **With TX0117 strains** CFU/g of vegetations (mean log10 +/- SD) after 3 days of treatment **With TX0117c strains** mean log10 reduction after 3 days of treatment | **With TX0117 strains** Cefazolin: 5.5 +/- 2.0, Dpatomycine 0.2 +/- 0.4, Nafcillin 2.0 +/- 2.9 (cefazolin versus daptomycin, $p< 0.0001$; cefazolin versus nafcillin, $p< 0.005$; daptomycin versus nafcillin, $p = 0.053$). **With TX0117c strains** Nafcillin 1.4 log10 Cefazolin 5.5 ($p=0.0001$), | - | NA |

1GC: 1st Generation Cephalosporin; 2GC: 2nd Generation Cephalosporin
*At the time of sacrifice, right atrial blood cultures

**Table 3.** Characteristics and results of studies comparing efficacy of cefazolin versus anti-staphylococcal penicillins in the treatment of MSSA bacteremia.

| | Study Design | Mechanisms to control bias | Antibiotics (dosing) | Number of patients | Severity of illness | Deep seated infections[¶] | Duration of bacteremia in days (mean (SD) or median [IQR]) | Metastatic infection | Source control[§] |
|---|---|---|---|---|---|---|---|---|---|
| **Paul et al. 2011** 1988 – 1994 & 1999-2007 Petah Tikva, Israel | Single center Retrospective cohort | Multivariate Logistic regression | CFZ | N = 72 | NA | 19% | NA | NA | NA |
| | | | Cloxacillin | N = 281 | | | | | |
| **Lee et al. 2011** 2004 – 2009 Seoul, South Korea | Single center Retrospective cohort | Propensity score | CFZ | N = 49 N' = 41 | Classified as ultimately or rapidly fatal according to McCabe score (%): 66 | 32%* | NA | 17% | 29% |
| | | | Nafcillin | N = 84 N' = 41 | Classified as ultimately or rapidly fatal according to McCabe score (%): 73 | 55%* | | 15% | 27% |
| **Li et al. 2014** 2008 – 2012 San Antonio, Tx, USA | Multicenter Retrospective cohort | Multivariate Logistic regression | CFZ (6g/day) | N = 59 | ICU admission (%): 7 Pitt Bacteremia Score (median, IQR): 0 [0-1] | 59% | 4 [2-6] | 34% | 56% |
| | | | Oxacillin (12g/day) | N = 34 | ICU admission (%): 18 Pitt Bacteremia Score (median, IQR): 0 [0-1] | 76% | 4 [3-7] | 35% | 50% |
| **Bai et al. 2015** 2007 – 2010 Toronto, Canada | Multicenter Retrospective cohort | Propensity score | CFZ (3g/day) | N = 105 N' = 90 | ICU admission (%): 10 | 32% | NA | NA | 63% |
| | | | Cloxacillin (12g/day) | N = 249 N' = 90 | ICU admission (%): 18 | 41% | | | 58% |
| **Rao et al. 2015** 2010 – 2013 Chicago, Il, USA | Multicenter Retrospective cohort | Multivariate Logistic regression | CFZ (4g/day) | N = 103 | ICU admission (%): 42 Modified-APACHE score (mean, SD): 13 (6.3) | 31% | 3 [2-4] | 29%* | 77%* |
| | | | Oxacillin (12g/day) | N = 58 | ICU admission (%): 33 Modified-APACHE score (mean, SD): 10.3 (5.8) | 35% | 3 [2-4] | 19% | 52%* |
| **Pollet et al. 2016** 2008 – 2013 San Francisco, Ca, USA | Single center Retrospective cohort | Propensity score | CFZ | N = 70 | ICU admission (%): 13 | 14% | 1.3 (0.8) | NA | NA |
| | | | Nafcillin | N = 30 | ICU admission (%): 27 | 30% | 1.7 (1.4) | | |
| **McDanel et al. 2017** 2003 – 2010 USA | Multicenter Retrospective cohort | - | CFZ | N = 1163 | ICU admission (%): 15 * APACHE III Score >34 (%): 56 | 41% | NA | NA | NA |
| | | | Nafcillin/Oxacillin | N = 2004 | ICU admission (%): 19 * APACHE III Score >34 (%): 52 | 43% | | | |

**Table 3 (continued).**

| | Antibiotics (dosing) | Number of patients | Efficacy Outcomes | Results | Results |
|---|---|---|---|---|---|
| **Paul et al. 2011** <br> **1988 – 1994 &** <br> **1999-2007** <br> **Petah Tikva, Israel** | CFZ | N = 72 | Day 90 mortality | 40% | Cefazolin aOR mortality <br> 0.91 [0.47–1.77] |
| | Cloxacillin | N = 281 | | 32% | |
| **Lee et al. 2011** <br> **2004 – 2009** <br> **Seoul, South Korea** | CFZ | N' = 41 | D90 Treatment failure (Change in antibiotic regimen, clinical failure, relapse or death) | 15% | Cefazolin aOR treatment failure <br> 1.6 [0.5 – 5.4] |
| | Nafcillin | N' = 41 | | 15% | |
| **Li et al. 2014** <br> **2008 – 2012** <br> **San Antonio, Tx, USA** | CFZ | N = 59 | D90 Treatment failure (persistent bacteremia, progression of infection, relapse or death) | 47% | - |
| | Oxacilline | N = 34 | | 24% | |
| **Bai et al. 2015** <br> **2007 – 2010** <br> **Toronto, Canada** | CFZ | N' = 90 | Day 90 mortality <br> Day 90 relapse | 20% vs 30% <br> 6% vs 2% | Cefazolin HR mortality <br> 0.58 [0.31 – 1.08] |
| | Cloxacillin | N' = 90 | | | |
| **Rao et al. 2015** <br> **2010 – 2013** <br> **Chicago, Il, USA** | CFZ | N = 103 | In-hospital mortality <br> Treatment failure | 1% vs 5% <br> 6% vs 12% | Oxacillin aOR treatment failure <br> 3.76 [0.98 – 14.4] |
| | Oxacillin | N = 58 | | | |
| **Pollet et al. 2016** <br> **2008 – 2013** <br> **San Francisco, Ca, USA** | CFZ | N = 70 | Day 90 mortality | 7% | Cefazolin aOR mortality <br> 0.40 [0.09 – 1.74] |
| | Nafcillin | N = 30 | | 17% | |
| **McDanel et al. 2017** <br> **2003-2010** <br> **USA** | CFZ | N = 1163 | Day 90 mortality <br> Day 90 recurrence | 25% vs 20%* <br> 20% vs 28% | Cefazolin aHR mortality <br> 0.77 [0.66 – 0.90] <br> Cefazolin aHR recurrence <br> 1.13 [0.94 – 1.36] |
| | Nafcillin/Oxacillin | N = 2004 | | | |

*$p<0.05$; aOR= adjusted odds ratio, CFZ=Cefazolin, HR= Hazard Ratio, MSSA= Methicillin-Susceptible *Staphylococcus aureus*, N'=number of included patients in the propensity score matched analyses,
¶ **Deep-seated infections: Endocarditis, Bone or joint infection, Device related infection, Deep-seated abcess, Pneumonia**
§ **Source control: Catheter removal, device explantation, surgical management of abcess**

**Table 4.** Characteristics and results of studies assessing safety of cefazolin versus anti-staphylococcal penicillins use in the treatment of MSSA bacteremia.

| | Study Design | Antibiotics (dosing) | Dosing | Median duration | Number of patients | Criteria | Results |
|---|---|---|---|---|---|---|---|
| **Lee 2011** 2004 – 2009 Seoul, South Korea | Single center Retrospective cohort | CFZ | NA | 17 [10-18] | N = 49 N' = 41 | Discontinuation AE | 0%* |
| | | Nafcillin | NA | 15 [10-25] | N = 84 N' = 41 | | 17%* |
| **Youngster 2014** 2007 – 2011 Boston, Ma, USA | Retrospective cohort | CFZ | 6g/day | NA | N = 119 | Premature ATB discontinuation Rash Nephrotoxicity Hepatotoxicity | 7% vs 34% 4% vs 14%* 3% vs 11%* 2% vs 8%* |
| | | Nafcillin | 8g/day | NA | N = 366 | | |
| **Li 2014** 2008 – 2012 San Antonio, Tx, USA | Multicenter Retrospective cohort | CFZ | 6g/day | 39 [28-44] | N = 59 | All AE Rash Elevated transaminases Elevated serum creatinine | 3% vs 30% * 2% vs 3% 0% vs 18%* 0 vs 3% |
| | | Oxacillin | 12g/day | 31 [21-42] | N = 34 | | |
| **Rao 2015** 2010 – 2013 Chicago, Il, USA | Multicenter Retrospective cohort | CFZ | 4g/day | 29 [15-42] | N = 103 | Any AE Rash Nephrotoxicity Hepatotoxicity | 4% vs 8% 0% vs 3% 0% vs 1% 0% vs 2% |
| | | Oxacillin | 12g/day | 32.5 [15-43] | N = 58 | | |
| **Flynt 2017** 2010 – 2013 Dertoit, MI, USA | Multicenter Retrospective cohort | CFZ | NA | NA | N = 68 | Acute Kidney Injury | 13%* |
| | | Naficillin | NA | NA | N = 81 | | 32%* |

*p<0.05
AE= adverse event, ASPs= anti-staphylococcal penicillins, ATB= Antibiotics, CFZ= Cefazolin, MSSA= methicillin-susceptible *Staphylococcus aureus*, NA= not available.